\documentclass{article}
\usepackage{spconf,amsmath,graphicx}

\usepackage{enumitem}
\setlist{nosep, leftmargin=14pt}

\usepackage{mwe} % to get dummy images

\usepackage[algo2e]{algorithm2e} 
\usepackage{algorithmicx}
\usepackage{caption}

\usepackage{booktabs}

\usepackage[table]{xcolor} %For table row-coloring

\usepackage{multirow}

\usepackage{algpseudocode}
\usepackage{algorithm}

\usepackage{array} % to control column spacing in matrix of 

\usepackage{pifont}

\title{
LEARNER: Contrastive Pretraining for Learning Fine-Grained Patient Progression from Coarse Inter-Patient Labels
}

\name{
\begin{tabular}{c}
Jana Armouti$^{\star}$ \qquad
Nikhil Madaan$^{\star}$ \qquad
Rohan Panda$^{\star}$ \qquad
Tom Fox$^{\dagger}$ \qquad
Laura Hutchins$^{\dagger}$ \qquad\\
Amita Krishnan$^{\dagger}$ \qquad
Ricardo Rodriguez$^{\ddagger}$ \qquad
Bennett DeBoisblanc$^{\dagger}$ \qquad
Deva Ramanan$^{\star}$ \qquad\\
John Galeotti$^{\star}$ \qquad
Gautam Rajendrakumar Gare$^{\star}$ 
\end{tabular}
}

\address{
$^{\star}$ Carnegie Mellon University, Pittsburgh, USA \\
$^{\dagger}$ LSUHSC Internal Medicine, New Orleans, USA \\
$^{\ddagger}$ Cosmetic Surgery Facility LLC, Baltimore, MD 21093, USA
}

\begin{document}

\maketitle

\begin{abstract}
Predicting whether a treatment leads to meaningful improvement is a central challenge in personalized medicine, particularly when disease progression manifests as subtle visual changes over time. While data-driven deep learning (DL) offers a promising route to automate such predictions, acquiring large-scale longitudinal data for each individual patient remains impractical. To address this limitation, we explore whether inter-patient variability can serve as a proxy for learning intra-patient progression. We propose \textbf{LEARNER}, a contrastive pretraining framework that leverages coarsely labeled inter-patient data to learn fine-grained, patient-specific representations. Using lung ultrasound (LUS) and brain MRI datasets, we demonstrate that contrastive objectives trained on coarse inter-patient differences enable models to capture subtle intra-patient changes associated with treatment response. Across both modalities, our approach improves downstream classification accuracy and F1-score compared to standard MSE pretraining, highlighting the potential of inter-patient contrastive learning for individualized outcome prediction.
\end{abstract}


\section{Introduction}
\label{sec:intro}

\begin{figure*}[ht]
    \centering
    \includegraphics[width = \linewidth, height=3.3cm]{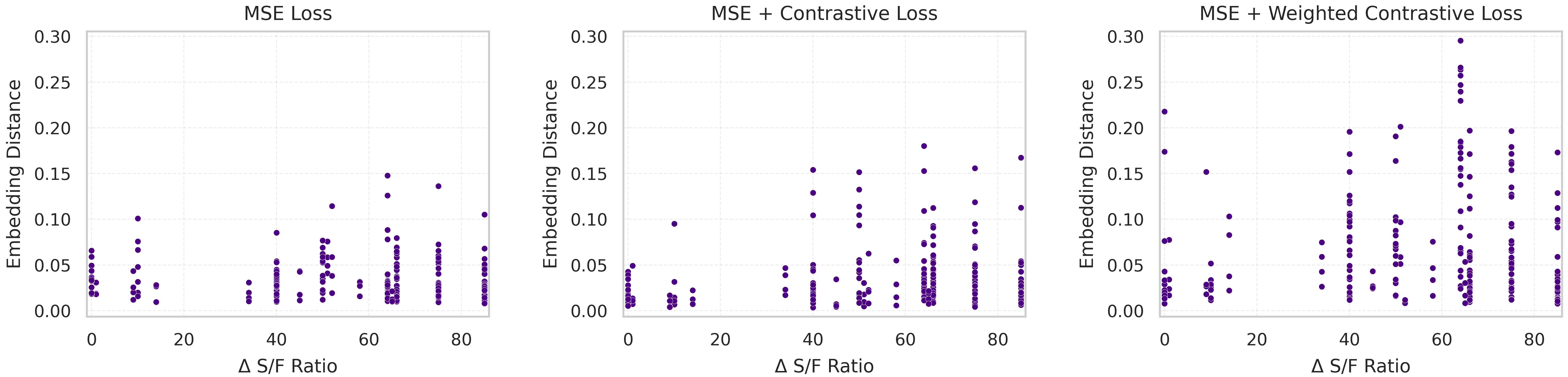}
   \caption{
   Comparison of embedding distance versus the actual change in physiological health score ($\Delta$S/F ratio) between day-1 and day-2 scans of individual patients. Each point represents a scan pair from the same patient, where the $x$-axis is the absolute difference in ground-truth S/F ratio between the two scans (larger values indicate greater clinical improvement or decline), and the $y$-axis (Embedding Distance) is the cosine distance between their latent embeddings. With MSE pretraining, many pairs with large S/F ratio differences still have small embedding distances (points clustered near the bottom-right), indicating that the representation underestimates clinical change. Adding contrastive pretraining shifts pairs with larger label differences upward and yields a more spread-out, approximately increasing pattern, with the weighted contrastive variant producing the strongest separation. This behavior suggests that contrastive pretraining aligns embedding distances more faithfully with clinical differences and improves sensitivity to subtle physiological progression.
   %DEVA: Need to provide more context to ground the figure; I frankly don't understand it. What are the labels and what do the colors mean? What does a "label difference of 100" mean? Perhaps most importantly, why is it better to for embedding distributions to be more spread out?
   % \john{With MSE loss the points are predominantly clustered in the bottom right corner, whereas our contrastive loss shifts some of the points to the left, better aligning the points with the diagonal.}
    }
    \label{fig:embeddingDist}
\end{figure*}

% Automated patient monitoring is critical for timely clinical decision-making, especially when clinical resources are constrained, as during the COVID-19 pandemic when clinicians were forced to make life-altering decisions about access to ventilators and oxygen~\cite{shortBosa2022Responseunprepared}.  %DEVA: first sentence is akward; didn't such decisions need to be made before COVID?
% For many hospitalized patients, response to treatment is tracked using imaging modalities such as lung ultrasound (LUS) or brain MRI. However, detecting meaningful treatment effects from these images is challenging: disease progression is gradual and patient-specific, and subtle temporal differences between scans are often obscured by strong patient-level correlations~\cite{shortRoshankhah2021InvestigatingImagesa}. As a result, models can easily learn to recognize patient identity while remaining insensitive to fine-grained intra-patient change.
Automated patient monitoring is crucial for responsive clinical decision-making, particularly during crises such as COVID-19, when clinicians faced resource allocation challenges for ventilators and oxygen~\cite{shortBosa2022Responseunprepared}. Detecting treatment effects from imaging modalities like lung ultrasound (LUS) is challenging due to gradual and patient-specific progression. Subtle temporal differences in scans are often obscured by patient-level correlations~\cite{shortRoshankhah2021InvestigatingImagesa}, limiting model sensitivity to intra-patient change.

Deep learning (DL) offers a natural route to automating longitudinal assessment, but state-of-the-art models typically require large, densely sampled cohorts with many labeled scans per patient, data that are difficult and expensive to collect in real-world settings. This motivates the question: can we leverage abundant inter-patient variability to learn representations that remain sensitive to subtle intra-patient progression, even for patients unseen during training?

% In this work, we address this question by proposing to train models on coarse inter-patient differences in order to capture fine-grained intra-patient changes. We introduce LEARNER, a contrastive pretraining framework that operates on batches of LUS or MRI clips from multiple patients and uses differences in their clinical scores as supervision. During pretraining, the model is encouraged to separate embeddings according to coarse label differences across patients, using a contrastive objective based on SpO2/FiO2 (S/F) ratio or Mini-Mental State Examination (MMSE) differences. We hypothesize that the resulting latent space will generalize to detect subtle within-patient changes over time

In this work, we address this question by proposing to train models on coarse inter-patient differences to induce sensitivity to fine-grained intra-patient changes. We introduce LEARNER, a contrastive pretraining framework that operates on batches of LUS or MRI clips from multiple patients and uses differences in their clinical scores as supervision. During pretraining, LEARNER explicitly encourages patient embeddings to separate according to coarse label disparities (S/F ratio for LUS and MMSE for MRI). We hypothesize that the resulting latent space will generalize to detect subtle within-patient changes over time.
% During pretraining, LEARNER explicitly encourages patient embeddings to separate according to coarse label disparities, using a contrastive objective based on SpO\textsubscript{2}/FiO\textsubscript{2} (S/F) ratio differences for LUS and Mini-Mental State Examination (MMSE) differences for MRI. We hypothesize that the resulting latent space will generalize to detect subtle within-patient changes over time.

% We evaluate LEARNER on two longitudinal datasets: (a) an in-house LUS dataset, where the goal is to track hypoxemia (low blood oxygen levels) via changes in S/F ratio across successive scans, and (b) the public ADNI Alzheimer's Disease MRI dataset, where the goal is to track cognitive decline via changes in MMSE. 
We evaluate LEARNER on two longitudinal datasets: an in-house LUS cohort for tracking hypoxemia and the ADNI MRI cohort for tracking cognitive decline. Across both, we find that contrastive pretraining on inter-patient label differences yields representations that better capture intra-patient progression (see Figure~\ref{fig:embeddingDist}), improving downstream prediction of treatment response and disease trajectory.

% \section{Related Work}
\textbf{Related Work.}
Contrastive learning~\cite{chen2020simple} has been widely studied in computer vision and medical imaging~\cite{zhang2020contrastive,wang2020contrastive,xue2021modality,gare2022weakly}. Prior work has primarily focused on improving diagnostic classification (e.g., COVID-19 versus healthy or other pneumonia), few-shot COVID-19 prediction \cite{zhang2020contrastive,wang2020contrastive}, or leveraging multimodal pairs of images and text \cite{xue2021modality}. Recently, \cite{gare2022weakly}  demonstrated a weakly supervised training method that applied contrastive learning to predict COVID-19 severity using noisy labels. In contrast, our goal is to learn embeddings that reflect clinically significant progression over time.

\vspace{-1.0em}
\section{Methodology}
\label{sec:method}
% We consider two related prediction tasks on longitudinal imaging data. First, we train our models on a regression task, where the goal is to predict a clinical score from a single scan (S/F ratio for LUS, MMSE for ADNI). Second, we evaluate on a three-way health-change classification task: given a pair of sequential scans from the same patient, the model predicts whether the patient's condition has improved, stayed the same, or deteriorated.
We address the challenge of modeling patient progression by decomposing it into two related prediction tasks using longitudinal imaging data. First, we pretrain our models on a single-scan regression task, where the goal is to predict a clinical score (S/F ratio for LUS, MMSE for ADNI). We then formulate the prognostic objective as a three-way health-change classification task; given a pair of sequential scans from the same patient, the model predicts whether the patient’s condition has improved, remained stable, or deteriorated.

\vspace{-1em}
\subsection{Datasets}
\label{sec:dataset}

% \subsubsection{LUS}
\textbf{LUS dataset.}
% We use an in-house lung ultrasound (LUS) dataset \cite{gare2022learning} consisting of linear probe B-mode videos collected from \textbf{221 patients} (a total of \textbf{1001 videos}). Each patient has multiple (at least two) longitudinal ultrasound scans of both left and right lung regions, acquired at imaging depths of 4-6\,cm under different scan settings using a Sonosite X-Porte ultrasound machine (IRB approval number 1509\footnote{The private LUS dataset, collected under IRB protocol number 1509 titled \textit{“Artificial Intelligence Interpretation of Lung Ultrasound Images”}, was de-identified prior to its transfer to CMU.}). 
We use an in-house LUS dataset \cite{gare2022learning} comprising 1001 linear probe B-mode videos from \textbf{221 patients}, each with $\geq 2$ longitudinal scans of both lungs, acquired at 4-6 cm depth on a Sonosite X-Porte machine 
(IRB 1509).
% (IRB 1509\footnote{Dataset collected under IRB protocol 1509, “Artificial Intelligence Interpretation of Lung Ultrasound Images”, and de-identified before transfer to CMU.}).

% We split the dataset at the patient level to avoid cross-patient contamination, ensuring that all videos from a given patient appear in only one split. This yields \textbf{160} patients for training, \textbf{44} for validation, and \textbf{17} for testing. The corresponding numbers of individual videos (used for the regression task) are listed in Table~\ref{tab:dataset}.
The data are split at the patient level to avoid leakage: 160 patients for training, 44 for validation, and 17 for testing. Only the test-set patients have sequential scans. Video counts for the regression task are listed in Table~\ref{tab:dataset}.

% For the health-change classification task, we form:
% Inter-patient pairs (train/val), created by randomly pairing videos from different patients;
% Intra-patient pairs (test), using scans acquired on different days.

For the health-change classification task, we form longitudinal pairs to capture progression over time. Specifically, we generate:
\begin{itemize}[leftmargin=1.2em]
    \item \textbf{Inter-patient pairs:} videos randomly paired across different patients (train and validation sets);
    \item \textbf{Intra-patient pairs:} videos from the same patient acquired on different days (test set).
\end{itemize}
% This results in \textbf{2103}, \textbf{582}, and \textbf{236} paired datapoints for the train, validation, and test splits for the three-way classification.
This produces 2103, 582, and 236 paired samples for train, validation, and test splits for the three-way classification task.

\textbf{S/F Ratio:} 
% The S/F ratio represents the ratio between measured blood oxyhemoglobin saturation (S) and the fraction of inspired oxygen (F). The lower this ratio, the more deranged the lung function is. S/F is a standardized measurement used to assess lung function in research and at bedside. We categorize the S/F ratio into the following 4 ranges: [$>$ 430, 275$-$430, 180$-$275, $<$ 180], as these ranges reflect the maximum the medical devices can support lungs in oxygenating blood. Changes in the S/F category are used as labels for health changes.
The S/F ratio (blood oxygen saturation / inspired oxygen fraction) is a standard measure of lung function; lower values indicate worse oxygenation. We discretize it into four clinically relevant ranges: [$>$ 430, 275$-$430, 180$-$275, $<$ 180]. Changes in these categories define the health-change labels.

\begin{figure*}[t]
\centering
% ---------------- LEFT: ALGORITHM ----------------
\begin{minipage}[t]{0.8\textwidth}
\RestyleAlgo{ruled}
\vspace{-1em}
\begin{algorithm}[H]   % <-- NON-FLOATING VERSION
\setlength{\algomargin}{0.8em}
\SetAlgoLined
\textbf{Require:} 
$\theta_V$, $\theta_M$ : parameters of the video network and MLP head.

\textbf{Require:} 
Pre-trained weights for encoder parameters $\theta_V$.

\textbf{Require:} 
{
$D:$ training set
\While{not done} {
    \For{each batch $\mathcal{B} = \{(x_i, HS_i)\}_{i=1}^{B}$ from $D$}{
        Do a forward pass through the video encoder, i.e. $U = \theta_V(\mathcal{B})$.
        
        Compute $d_{ij} = \left \| HS_i - HS_j \right \|$ for all pairs of datapoints in the current batch.

        Compute predictions by doing a forward pass through the MLP head, i.e. $Y = \theta_M(U)$.

        For each $i$, sort $\{d_{ij}\}_j$ and select the first $\frac{B}{2}$ indices as positive points and the last $\frac{B}{2}$ indices as negative points.

        Compute the MSE loss (Eq.~\ref{eq:l2loss}) and the contrastive loss (Eq.~\ref{eq:clLoss}) to obtain the final loss.

        Update $\theta_V$ and $\theta_M$ using the combined loss function: $
        \theta \leftarrow \theta - \eta\nabla_{\theta}\big(\lambda \mathcal{L}_{\mathrm{MSE}} + (1-\lambda) \mathcal{L}_{\mathrm{CL}}\big)
        $, where $\theta = \{\theta_V,\theta_M\}$.
    }
}
}
\caption{Contrastive Learning based Pre-training}
\label{alg:proposed_model}
\end{algorithm}
\end{minipage}
\hfill
% ---------------- RIGHT: TABLES ----------------
\begin{minipage}[t]{0.19\textwidth}
\captionof{table}{Dataset splits and patient counts \& individual scans/videos used for regression.}
\label{tab:dataset}
\resizebox{\linewidth}{!}{
\begin{tabular}{l|cccc}
\toprule
\multirow{2}{*}{\textbf{Split}} &
\multicolumn{2}{c}{\textbf{\# of Datapoints}} &
\multicolumn{2}{c}{\textbf{\# of Patients}} \\
 & LUS & ADNI & LUS & ADNI \\
\midrule
Train & 701 & 960 & 160 & 320 \\
Val   & 194 & 480 & 44  & 160 \\
Test  & 106 & 416 & 17  & 159 \\
\bottomrule
\end{tabular}
}
\vspace{0.5em}
\captionof{table}{Training hyperparameters.}
\label{tab:hyperparams}
\resizebox{\linewidth}{!}{
\begin{tabular}{l|l}
\toprule
\textbf{Hyperparameter} & \textbf{Value} \\
\midrule
Batch Size  & 16 \\
Initial LR  & 0.001 \\
Optimizer   & Adam \\
Epochs      & 100 \\
Scheduler   & Cosine Annealing \\
\bottomrule
\end{tabular}
}
\end{minipage}
\end{figure*}

% \begin{table}[h]
%  \begin{minipage}{.45\linewidth}
%     \begin{center}
%     \caption{Dataset splits (individual scans/videos used for regression) and unique patients in each split.}\label{tab:dataset}
%     \resizebox{1.0\textwidth}{!}{
%     \begin{tabular}{l|cccc}
%     \toprule
%     \multirow{2}{*}{\textbf{Split}} & \multicolumn{2}{c}{\textbf{\# of Datapoints}} & \multicolumn{2}{c}{\textbf{\# of Patients}} & \\ 
%      & LSU & ADNI & LSU & ADNI \\ 
%     \midrule
%     Train           & 701   & 960  & 160 & 320              \\ 
%     Val             & 194   & 480  & 44 &  160              \\ 
%     Test            & 106   & 416  & 17 &  159              \\  
%     \bottomrule
%     \end{tabular}
%     }
%     \end{center}
% \end{minipage}
% \hspace{1em}
%     \begin{minipage}{.45\linewidth}
%     \begin{center}
%     \caption{Train params.}
%     \label{tab:hyperparams}
%     \resizebox{0.8\textwidth}{!}{
%     \begin{tabular}{l|l}
%     \toprule
%     \textbf{Hyperparam} & \textbf{Value} \\
%     \midrule
%     BatchSize               & 16                        \\ 
%     Initial LR              & 0.001                    \\ 
%     Optimizer               & Adam                     \\ 
%     Epochs                  & 100                      \\ 
%     Scheduler               & CosineAnnealing          \\
%     \bottomrule
%     \end{tabular}
%     }
%     \end{center}
%     \end{minipage}%
% \end{table}

% \RestyleAlgo{ruled}
% \input{algo/proposed_algo.tex}

% \subsubsection{ADNI} 
\textbf{ADNI dataset} 
% The Alzheimer's Disease Neuroimaging Initiative (ADNI) \cite{shortWeiner2010ThePlans} is a large longitudinal study designed to identify early markers of Alzheimer's disease progression. It includes data from over 1,500 participants (Alzheimer's disease, mild cognitive impairment (MCI), and healthy controls) collected across more than 50 sites in the U.S. and Canada. The dataset encompasses MRI, PET, genetic and clinical data, as well as repeated cognitive assessments. For our analysis, we focus on structural MRI and the Mini-Mental State Examination (MMSE), a widely used mental status test that is tracked over time in ADNI. We use the ADNI phase-1 cohort (ADNI-1), which includes \textbf{639} patients with up to three consecutive MRI scans acquired at six-month intervals. We select patients with at least two visits and split them into training, validation, and test sets using a 2:1:1 patient-level ratio, resulting in the scan and patient counts summarized in Table \ref{tab:dataset}. Similar to the LUS setting, we use the individual MRI scans and their MMSE scores as datapoints for the regression task (predicting MMSE from a single scan). For the health-change classification task, we form sequential scan pairs for each patient, and assign labels based on the change in MMSE score between visits.
The Alzheimer’s Disease Neuroimaging Initiative (ADNI) \cite{shortWeiner2010ThePlans} is a large, multi-site longitudinal study that tracks cognitive and imaging biomarkers across more than 1,500 participants. For this work, we focus on structural MRI and the \textbf{Mini-Mental State Examination (MMSE)}, a widely used cognitive score measured repeatedly in ADNI.

We use the ADNI-1 cohort, which contains \textbf{639} patients with up to three MRI scans acquired at six-month intervals. We select patients with at least two visits and split them at the patient level in a 2:1:1 ratio, yielding the scan and patient counts shown in Table~\ref{tab:dataset}. Individual MRI scans paired with their MMSE scores form the datapoints for the regression task. For the health-change classification task, we construct sequential scan pairs for each patient and assign labels based on the change in MMSE between visits.

% \section{Methodology}

\begin{figure}[h]
    \centering
    \includegraphics[width = \linewidth]{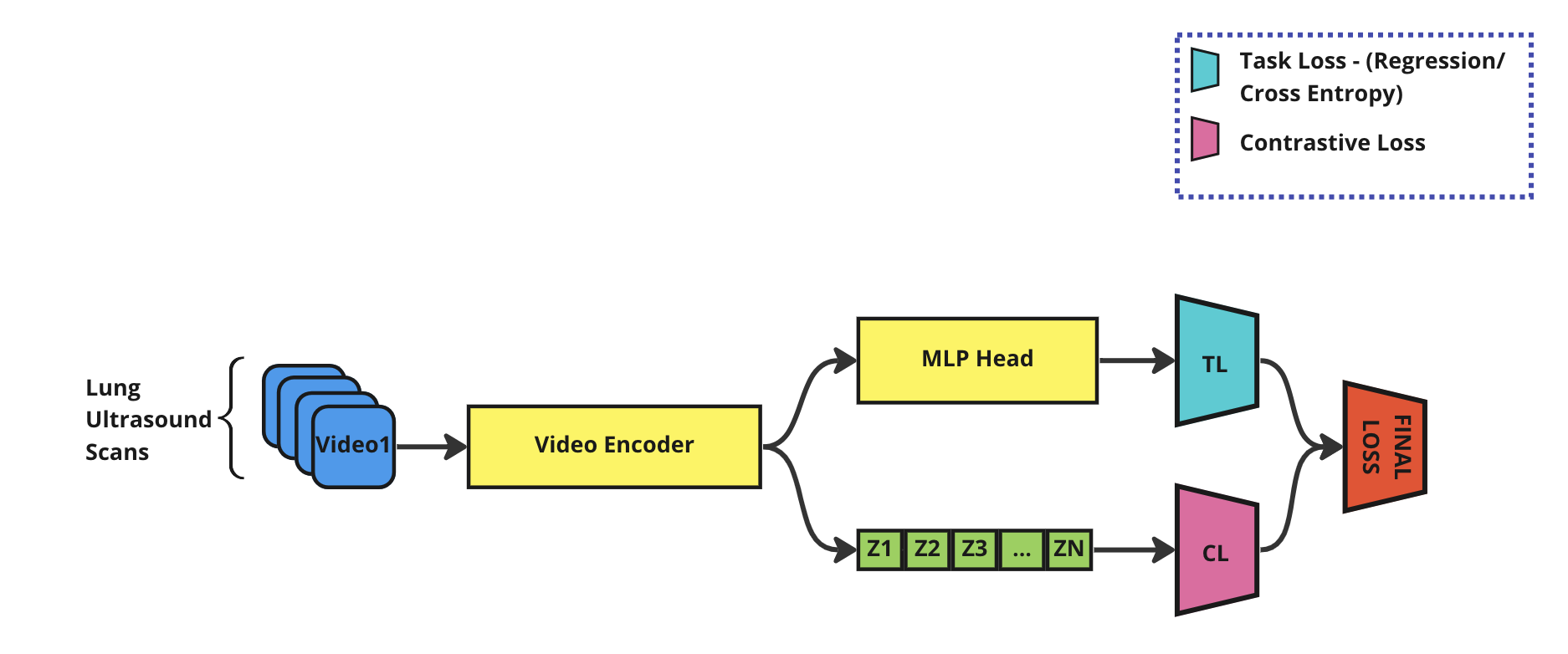}
    \caption{Proposed Model Architecture}
    \label{fig:model}
\end{figure}

\vspace{-1.0em}
\subsection{Model Architecture}
\vspace{-0.5em}
% We use the Temporal Shift Module (TSM) video network \cite{lin2019tsm} with the ResNet-18 backbone. TSM model aims to provide the benefits and competitive performance of a 3D CNN while enjoying the complexity of a 2D CNN. It infuses temporal information into every 2D CNN resnet block, by shifting certain channels from the previous and next time-frame. Refer to \cite{lin2019tsm} for details. 
We use the Temporal Shift Module (TSM) network \cite{lin2019tsm} with a ResNet-18 backbone. TSM provides the temporal modeling benefits of a 3D CNN while retaining the computational efficiency of a 2D CNN by shifting a subset of feature channels forward and backward in time within each ResNet block.
\footnote{See \cite{lin2019tsm} for details.}
% See \cite{lin2019tsm} for details.

\vspace{-1em}
\subsection{Pre-training task:}
\vspace{-0.5em}
% We pre-train the encoder by training the model to predict the patient health-score [$HS$] (S/F or MMSE) by formulating it as a regression task i.e. training a model to predict the S/F ratio (MMSE) for a given LUS (MRI). Our pre-training algorithm is described in Alg. \ref{alg:proposed_model}.
We pre-train the encoder by training the model to predict the patient health score [$HS$] (S/F for LUS, MMSE for ADNI) from a single scan, formulated as a standard regression task. The full pretraining procedure is summarized in Alg.~\ref{alg:proposed_model}.

\subsubsection{Loss functions:}

\textbf{MSE Loss ($MSE$):} We first supervise the model with a standard mean-squared error (MSE) loss between the predicted and ground-truth health scores:
\begin{equation}
\mathcal{L}_{\mathrm{MSE}} 
= \sum_{i=1}^{B} \left\| HS_i - \hat{HS}_i \right\|^2 ,
\label{eq:l2loss}
\end{equation}
where $B$ is the batch size.

\vspace{-0.1em}
\textbf{Contrastive Loss ($CL$):} To leverage inter-patient variability, we introduce a batch-wise contrastive term defined over pairs of scans with different clinical scores. For a batch of embeddings $\{u_i\}$ with corresponding health scores $\{HS_i\}$, we define positives and negatives for each sample $i$ by ranking the other samples in the batch according to their absolute score difference $\left \| HS_i - HS_j \right \|$ and selecting the most similar and most dissimilar examples in score space (see Algorithm~\ref{alg:proposed_model} for details). Using cosine similarity $\mathrm{sim}(u_i,u_j)$ between embeddings, the resulting (unweighted) contrastive loss is:
\begin{equation}
\mathcal{L}_{\mathrm{CL}} 
= \sum_i \left( 
    \sum_{j \in i^{-}} \log \mathrm{sim}(u_i,u_j) 
  - \sum_{j \in i^{+}} \log \mathrm{sim}(u_i,u_j)
\right)
\label{eq:cl_unweighted}
\end{equation}
where $i^{+}$ and $i^{-}$ denote the positive and negative sets for anchor $i$. This encourages embeddings of clinically dissimilar patients (large score differences) to be far apart, and embeddings of clinically similar patients to be close. %DEVA: It seems like this is trying to increase the variance of the features; might be useful to cite a method that does this (eg, VICReg, ICLR22)

\textbf{Weighted Contrastive Loss ($wCL$):} In practice, we found it beneficial to modulate each contrastive term by the underlying clinical score difference (S/F or MMSE) of the two sequential scans to determine the weight of the corresponding term in the loss function:
\[
w_{ij} = d_{ij} + \varepsilon = \left \| HS_i - HS_j \right \| + \varepsilon,
\]
with a small $\varepsilon > 0$ for numerical stability, and use the following weighted contrastive loss:
% \begin{equation}
% \begin{aligned}
% \mathcal{L}_{\mathrm{wCL}} 
% ={}& \sum_{i} \sum_{j \in i^{-}} 
%          \log\big( \mathrm{sim}(u_i, u_j)\, w_{ij} \big) \\
%    & {}- \sum_{i} \sum_{j \in i^{+}} 
%          \frac{\log \mathrm{sim}(u_i, u_j)}{w_{ij}}
% \end{aligned}
% \label{eq:clLoss}
% \end{equation}
\begin{equation}
\begin{aligned}
\mathcal{L}_{\mathrm{wCL}} 
={}& \sum_{i} \sum_{j \in i^{-}} 
         \log\big( \mathrm{sim}(u_i, u_j)\, w_{ij} \big) \\
   & {}- \sum_{i} \sum_{j \in i^{+}} 
         \frac{\log \mathrm{sim}(u_i, u_j)}{w_{ij}}
\end{aligned}
\label{eq:clLoss}
\end{equation}
Pairs with large label differences receive larger weights in the negative term and smaller weights in the positive term, further separating clinically dissimilar patients in the embedding space while pulling together clinically similar ones. Note that the unweighted contrastive loss in Eq.~\eqref{eq:cl_unweighted} is recovered by setting $w_{ij}$ in Eq.~\eqref{eq:clLoss} to $1$ for all pairs.

\textbf{Total pretraining loss:}
Depending on the experiment, we use one of the following objectives:
\[
\mathcal{L}_{\text{pretrain}} =
\begin{cases}
\mathcal{L}_{\mathrm{MSE}}, & \text{(MSE)} \\
\lambda\mathcal{L}_{\mathrm{MSE}} + (1-\lambda) \mathcal{L}_{\mathrm{CL}}, & \text{(MSE+CL)} \\
\lambda\mathcal{L}_{\mathrm{MSE}} + (1-\lambda) \mathcal{L}_{\mathrm{wCL}}, & \text{(MSE+wCL)},
\end{cases}
\]
with weighting factor $\lambda$ (set to $
0.9$ in our experiments).

% \vspace{-0.45cm}
\subsection{Downstream Classification task:}
After pretraining, we use the encoder from the pretrained model as a feature extractor for the health-change classification task.
For two sequential scans (from different patients for train, and same patient for test), we obtain embeddings $u_t$ and $u_{t+1}$. We combine them using the embedding difference $d_t = u_{t+1} - u_t$, and feed it into a three-way MLP classifier trained with cross-entropy loss to predict whether the health score ($HS$) has improved, remained the same, or deteriorated between the two scans.

\vspace{-1em}
\section{Results}
\vspace{-1em}

\begin{table}[h]
\caption{Classification results using MSE and CL based pre-training. We observe that supplementing the MSE loss with CL significantly improves the performance, \emph{highlighting the benefit of inter-patient comparisons in improving same-patient predictions through contrastive learning.}
}
\label{tab:resCNN}
\centering
\resizebox{0.9\linewidth}{!}{
\begin{tabular}{l|cc|cc}
\hline
\multirow{2}{*}{\textbf{Loss}}    &  \multicolumn{2}{c}{\textbf{LUS Dataset}} & \multicolumn{2}{c}{\textbf{ADNI Dataset}} \\ 
\cmidrule{2-3}  \cmidrule{4-5} \\ [-\normalbaselineskip]
                        & \textbf{Accuracy} & \textbf{F1 Score}  & \textbf{Accuracy} & \textbf{F1 Score} \\
\hline
$MSE$           & 52.54           & 51.93           & 31.49           & 31.37   \\  
$MSE$ + $CL$    & 58.05           & 57.54           & 33.65           & 33.55   \\ 
$MSE$ + $wCL$   & \textbf{72.46}  & \textbf{71.25}  & \textbf{35.34}  & \textbf{34.53}   \\ \hline
\end{tabular}
}
\end{table}

% \begin{table}[ht]
\begin{table}[t]
\centering
\caption{
Ablation study on LUS dataset comparing backbone architectures (TSM, ViT, Swin), embedding-combination strategies (difference vs. concatenation), and pretraining objectives (MSE, MSE+CL, MSE+wCL). Results show that difference-based pairing and weighted contrastive supervision (MSE+wCL) consistently improve performance, with TSM and ViT emerging as the most data-efficient backbones.
}
\label{tab:appendix_ablations_full}
% \resizebox{1\linewidth}{!}{
% \resizebox{1\linewidth}{5.5cm}{
\resizebox{1\linewidth}{5.75cm}{
\begin{tabular}{llccccc}
\toprule
\textbf{Backbone} & \textbf{Combine} & \textbf{Loss} & \textbf{Classifier} & \textbf{Accuracy} & \textbf{F1} \\
\midrule
% 
% ------------------------- TSM ----------------------------
\multirow{12}{*}{\textbf{TSM}} 
& \multirow{6}{*}{Difference} 
  & MSE       & MLP Large & 55.51 & 55.33 \\
& & MSE+CL    & MLP Large & 54.66 & 54.36 \\
& & MSE+wCL   & MLP Large & 70.76 & 70.04 \\
& & MSE       & MLP       & 52.54 & 51.93 \\
& & MSE+CL    & MLP       & 58.05 & 57.54 \\
& & MSE+wCL   & MLP       & \textbf{72.46} & \textbf{71.25} \\
\cmidrule(lr){2-6}
& \multirow{6}{*}{Concatenate} 
  & MSE       & MLP Large & 50.85 & 50.48 \\
& & MSE+CL    & MLP Large & 53.81 & 52.31 \\
& & MSE+wCL   & MLP Large & 66.10 & 62.91 \\
& & MSE       & MLP       & 59.75 & 59.83 \\
& & MSE+CL    & MLP       & 52.12 & 50.47 \\
& & MSE+wCL   & MLP       & \textbf{68.64} & \textbf{65.95} \\
\midrule
% 
% ------------------------- SWIN ----------------------------
\multirow{12}{*}{\textbf{Swin}} 
& \multirow{6}{*}{Difference} 
  & MSE       & MLP Large & 27.97 & 21.28 \\
& & MSE+CL    & MLP Large & 30.51 & 14.26 \\
& & MSE+wCL   & MLP Large & \textbf{40.68} & \textbf{23.52} \\
& & MSE       & MLP       & 26.69 & 19.80 \\
& & MSE+CL    & MLP       & 30.51 & 14.26 \\
& & MSE+wCL   & MLP       & 30.51 & 14.26 \\
\cmidrule(lr){2-6}
& \multirow{6}{*}{Concatenate} 
  & MSE       & MLP Large & 28.81 & 12.89 \\
& & MSE+CL    & MLP Large & \textbf{40.68} & \textbf{23.52} \\
& & MSE+wCL   & MLP Large & \textbf{40.68} & \textbf{23.52} \\
& & MSE       & MLP       & 28.81 & 12.89 \\
& & MSE+CL    & MLP       & 30.51 & 14.26 \\
& & MSE+wCL   & MLP       & \textbf{40.68} & \textbf{23.52} \\
\midrule
% 
% ------------------------- VIT ----------------------------
\multirow{12}{*}{\textbf{ViT}} 
& \multirow{6}{*}{Difference} 
  & MSE       & MLP Large & 48.73 & 47.53 \\
& & MSE+CL    & MLP Large & 55.93 & 55.37 \\
& & MSE+wCL   & MLP Large & \textbf{72.88} & \textbf{72.77} \\
& & MSE       & MLP       & 46.19 & 46.12 \\
& & MSE+CL    & MLP       & 64.83 & 64.82 \\
& & MSE+wCL   & MLP       & 70.76 & 70.47 \\
\cmidrule(lr){2-6}
& \multirow{6}{*}{Concatenate} 
  & MSE       & MLP Large & 42.37 & 41.96 \\
& & MSE+CL    & MLP Large & 63.56 & 63.46 \\
& & MSE+wCL   & MLP Large & \textbf{66.53} & \textbf{65.44} \\
& & MSE       & MLP       & 47.03 & 46.32 \\
& & MSE+CL    & MLP       & 63.14 & 63.02 \\
& & MSE+wCL   & MLP       & 65.68 & 62.97 \\
\bottomrule
\end{tabular}
}
\end{table}

% \vspace{-3em}
\subsection{Performance}
% We first compare pretraining with MSE alone, MSE plus contrastive loss (MSE+CL), and MSE plus weighted contrastive loss (MSE+wCL) on both LUS and ADNI using the TSM backbone (Table~\ref{tab:resCNN}). Basic hyperparameters are shared across all models (Table~\ref{tab:hyperparams}), and data splits are patient-level (Table~\ref{tab:dataset}), ensuring that scans from a given patient appear in only one split. This approach guarantees a valid evaluation by preventing overlap between training and testing clips from the same patient. Classification metrics are computed on held-out intra-patient scan pairs, i.e., sequential scans from the same patient.
% 
% Adding a contrastive objective on inter-patient label differences improves downstream health-change classification over MSE pretraining alone on both datasets (MSE+CL vs. MSE). The weighted variant (MSE+wCL), which scales contrastive terms by the difference in clinical scores, shows the largest gains: on LUS, accuracy/F1 increase from 52.5/51.9 to 72.5/71.3, and on ADNI from 31.5/31.4 to 35.3/34.5. These results indicate that inter-patient contrastive pretraining substantially improves sensitivity to subtle within-patient health changes.
We compare three pretraining objectives: MSE, MSE+CL, and MSE+wCL, on both LUS and ADNI using the TSM backbone (Table~\ref{tab:resCNN}). All models share the same hyperparameters (Table~\ref{tab:hyperparams}) and patient-level splits (Table~\ref{tab:dataset}) to avoid cross-patient leakage. Classification metrics are computed on held-out sequential scan pairs from the same patient.

Adding contrastive supervision improves downstream health-change classification over MSE alone, with the weight\-ed variant (MSE+wCL) providing the largest gains. On LUS, accuracy/F1 improve from 52.5/51.9 to 72.5/71.3, and on ADNI from 31.5/31.4 to 35.3/34.5. These results show that inter-patient contrastive pretraining substantially enhances sensitivity to subtle within-patient changes.

\textbf{Embedding analysis.}
% To further analyze representation quality, we look at the relationship between embedding distance and change in clinical score (Fig.~\ref{fig:embeddingDist}). We plot the cosine distance between embeddings of day-1 and day-2 scans from the same patient against the absolute difference in S/F ratio. With MSE-only pretraining, many pairs with large S/F changes still lie at small embedding distances, indicating that the representation underestimates clinical change. In contrast, contrastively pretrained models, especially with weighted contrastive loss (Eq ~\ref{eq:clLoss}), produce a more spread-out, approximately increasing pattern, with larger label differences corresponding to larger embedding distances. This aligns with our goal of learning embeddings that reflect clinically meaningful progression.
To assess representation quality, we examine how embedding distance relates to clinical score change (Fig.~\ref{fig:embeddingDist}). With MSE-only pretraining, pairs with large S/F changes often remain close in embedding space, indicating underestimation of clinical progression. Contrastive pretraining, especially the weighted variant produces a clearer, increasing relationship, with larger S/F differences corresponding to larger embedding distances, aligning with our goal of learning clinically meaningful representations.

\vspace{-1em}
\subsection{Ablations}
% We carry out a series of ablations on the LUS dataset (Table \ref{tab:appendix_ablations_full}) to understand the effect of backbone, sequence combination strategy, and loss function. We compare a video CNN (TSM) with 2D transformer backbones (Swin \cite{Liu2021SwinWindows}, ViT\cite{Dosovitskiy2020AnScale}), and for each model we vary how sequential scans are combined (embedding difference vs. concatenation) and whether we use a small MLP (1 hidden layer) or a larger MLP (3 hidden layer) classifier. Across all backbones and combination strategies, adding the weighted contrastive term (MSE+wCL) consistently yields the best performance, often providing large improvement over only MSE regression. The Swin backbone underperforms both TSM and ViT, likely reflecting its higher data requirements and the limited dataset size. Overall, these results suggest that contrastive supervision is most effective when weighted by clinical score differences, and TSM and ViT with difference-based combination are strong, data-efficient choices for modeling longitudinal progression.
We conduct ablations on LUS dataset (Table~\ref{tab:appendix_ablations_full}). Comparing a video CNN (TSM) with 2D transformer backbones (Swin \cite{Liu2021SwinWindows}, ViT \cite{Dosovitskiy2020AnScale}), we find that Swin consistently underperforms, likely due to its higher data requirements and our modest dataset size. TSM and ViT perform similarly, with ViT achieving its best results using a larger MLP head (3 hidden layers) and TSM performing best with a smaller MLP (1 hidden layer). We also observe that embedding differences outperform concatenation when combining sequential scans.

Across all backbones and pairing strategies, our weighted contrastive loss (MSE+wCL) yields the strongest results, substantially improving over MSE-only regression. Overall, these findings indicate that contrastive supervision is most effective when weighted by clinical score differences, and that TSM and ViT with difference-based pairing are strong, data-efficient choices for modeling longitudinal progression.

% \section{Limitations}
% \todoGRG{1) Need Limitations section, 2) Need future extension 3) Update fig-2 with pre-training info 4) Clean up the pre-training exp 5) Fig-1 need quantitative numbers comparing the spread 6) Add to related works section  https://ieeexplore.ieee.org/document/10635225 }

\vspace{-0.5cm}
\section{Conclusion}
% \vspace{-0.1em}
% In this work, we introduced LEARNER, a contrastive pretraining framework that uses coarse inter-patient label differences to learn representations sensitive to fine-grained intra-patient progression. By combining regression on clinical scores with (weighted) contrastive objectives defined over inter-patient pairs, our method learns an embedding space in which distances more reflect clinically meaningful change. On an in-house lung ultrasound dataset (S/F ratio) and the public ADNI brain MRI dataset (MMSE), LEARNER improves three-way health-change classification over MSE-only pretraining, particularly when using the weighted contrastive loss. These results suggest that inter-patient contrastive learning is a promising strategy for individualized outcome prediction from limited longitudinal data, and motivate future work on extending this approach to larger cohorts and additional imaging modalities.
% 
% In this work, we introduced LEARNER, a contrastive pretraining framework that leverages coarse inter-patient label differences to learn representations sensitive to fine-grained intra-patient progression. By combining clinical-score regression with (weighted) contrastive objectives over inter-patient pairs, LEARNER produces an embedding space in which distances more faithfully reflect clinically meaningful change.
%
In this work, we introduced LEARNER, a contrastive pretraining framework that uses coarse inter-patient label differences to learn representations sensitive to fine-grained intra-patient progression. By combining clinical-score regression with (weighted) contrastive objectives over inter-patient pairs, LEARNER yields an embedding space whose distances more faithfully track clinically meaningful change.

Across an in-house LUS dataset (S/F ratio) and the ADNI MRI dataset (MMSE), LEARNER improves three-way health-change classification over MSE-only pretraining, with the \emph{weighted contrastive loss} providing the largest gains. These findings demonstrate that inter-patient contrastive learning is a promising approach for individualized outcome prediction from limited longitudinal data and motivate extending this framework to larger cohorts and additional imaging modalities.

\pagebreak
\section{Acknowledgments}
This work was supported in part by the Center for Machine Learning and Health (CMLH) Translational Fellowship at Carnegie Mellon University (CMU). The authors thank the clinicians at Louisiana State University (LSU) for their invaluable assistance with data collection and the CMU research team for their insightful contributions. All data used in this study were collected under IRB protocol number 1509, Artificial Intelligence Interpretation of Lung Ultrasound Images, and were fully de-identified prior to transfer to CMU.

% \pagebreak
%%%%%%%%% REFERENCES
% \bibliographystyle{splncs04}
\bibliographystyle{IEEEbib}
\bibliography{egbib,short_ref,references}

\end{document}